\documentclass[aps,prd,twocolumn,showpacs,groupedaddress,nofootinbib,amsmath]{revtex4}
\usepackage{graphicx}
\usepackage{epsfig}
\usepackage{multirow}

\begin{document}

\title{ Nature of the Roberge-Weiss transition end points in two-flavor lattice QCD with  Wilson quarks}

\author{Liang-Kai Wu}
\thanks{Corresponding author. Email address: wuliangkai@163.com}
\affiliation{Faculty of Science, Jiangsu University, Zhenjiang 212013, People¡¯s Republic of China}

\author{Xiang-Fei Meng}
\affiliation{National Supercomputer Center, Tianjin, 300457, People¡¯s Republic of China}

\date{\today}

\begin{abstract}
We make simulations with  2 flavor Wilson fermions to investigate the nature of the end points of Roberge-Weiss (RW) first order phase transition lines. The simulations are carried out at 9 values of  the hopping parameter $\kappa$ ranging from 0.155 to 0.198 on different lattice spatial volume. The Binder cumulants,  susceptibilities and reweighted distributions
of the imaginary part of Polyakov loop are employed to determine the nature of the end points of RW transition
lines. The simulations show that the RW end points are of  first order at the values of $\kappa$ in our simulations.
\end{abstract}


\maketitle

\section{INTRODUCTION}
\label{SectionIntro}

A full understanding of QCD phase diagram is of great importance theoretically and phenomenologically.
The QCD phase diagram addresses which forms of nuclear matter exist at different finite temperature and
baryon density, and whether there are {\it bona fide} phase transition separating them, thus it is essential for
relativistic heavy ion collision experiments and astrophysics. QCD is a strongly interacting theory on the
scales of a baryon mass and below, so non-perturbative calculations from the first principle are preferrable.
Despite that substantial progress has been made with Monte Carlo simulations of lattice QCD at zero baryon density,
the studies at nonzero baryon density are haunted by the ``sign'' problem, for example, see Ref.~\cite{Kogut:2007mz}.
  To date
many indirect methods have been proposed to circumvent the sign problem, overviews with
references to these methods can be found in Ref.~\cite{Kogut:2007mz,Schmidt:2006us}. One of these methods consists of simulating
QCD with the imaginary chemical potential for which the fermion determinant is
positive~\cite{deForcrand:2010he,D'Elia:2009tm,D'Elia:2009qz,Nagata:2011yf,Bonati:2010gi,D'Elia:2007ke,Wu:2006su,deForcrand:2006pv,deForcrand:2008vr}. Full information
can be obtained by using the imaginary chemical potential which allows for analytic continuation via truncated
polynomials.

 The phase structure of QCD with imaginary chemical potential  not only deserves detailed investigations in its own
right theoretically, but also has significant relevance to physics at zero or small real chemical potential\cite{D'Elia:2009qz,D'Elia:2007ke,Bonati:2010gi,Kouno:2009bm,Sakai:2009vb,deForcrand:2010he,Aarts:2010ky,Philipsen:2010rq,Bonati:2012pe}.
 QCD with imaginary chemical potential has a rich phase diagram as a function of  imaginary chemical potential and quark masses.

In this paper, we present a study of phase structure of QCD at fixed imaginary chemical potential $\theta = \mu_I/T = \pi$ for
$N_f=2$ QCD with Wilson quarks. The partition function including the imaginary chemical potential is
\begin{eqnarray}
Z(T,\mu_I)={\rm Tr}\biggl(e^{-\frac{1}{T}(H-i\mu_IN)}\biggr),
\end{eqnarray}
Roberge and Weiss made the essential work with the imaginary chemical potential~\cite{Roberge:1986mm}, they found that
the partition function of QCD with imaginary chemical potential has two important symmetries, reflection symmetry in $\mu=\mu_R+i\mu_I$ and
periodicity in imaginary chemical potential,
\begin{eqnarray} Z(T,\mu)=Z(T,-\mu),\end{eqnarray}
\begin{eqnarray}Z(\mu/T) = Z(\mu/T+i2\pi n /3). \end{eqnarray}
 The periodicity is smoothly realized
in the low temperature, strong-coupling regime, whereas in the high temperature, weak-coupling regime, it is realized in a non-analytic
way. At high temperature, the system undergoes  a first order transition (RW transition) at critical values
of the imaginary chemical potential $\mu_I/T =
(n+\frac{1}{2}){2\pi}/3$
~\cite{Roberge:1986mm,deForcrand:2002ci,Lombardo} between adjacent Z(3) sectors and these Z(3) sectors are characterized by the Polyakov loop.
Thus the picture for the $T-\theta$ phase diagram is that repeats with a periodicity the first order transition line in the high temperature regime which necessarily ends at an end point at some temperature $T_{RW}$ when the temperature is decreased sufficiently.

At these end points, there are evidence that the analytic continuation of deconfinement/chiral transition line
from real chemical potential to
imaginary chemical one meets the RW transition line.
Recent numerical studies show that the RW transition line end points are  triple points for small and heavy quark mass,
and second order end points for intermediate quark masses. So there exist two tricritical points which separate the first order
regime from the second one~\cite{deForcrand:2010he,D'Elia:2009qz,Bonati:2010gi}. Moreover, it is pointed out~\cite{deForcrand:2010he,Philipsen:2010rq,Bonati:2012pe} that the scaling behaviour at the tricritical points
may shape the the critical line for real chemical potential, and subsequently, the line for real chemical potential is
qualitatively consistent with the scenario suggested in Ref.~\cite{deForcrand:2006pv,deForcrand:2008vr} which show that the first order region shrinks with the increasing real chemical potential.

Most of studies of finite temperature QCD have been performed using
staggered fermion action or the improved
versions~\cite{Karsch,Cheng:2006qk,Cheng:2006aj,Bernard:2004je,Aoki:2006br,Bernard:2007nm,Aoki:2005vt},
 staggered fermion approach and Wilson
fermion approach have their own advantages and disadvantages, for example, see Ref.~\cite{Heller:2006ub}. The KS fermion formalism preserves the U(1)
chiral symmetry, whereas it needs a fourth root trick for one flavour which might lead to locality problem \cite{Bunk} and phase ambiguities~\cite{Golterman}. On the contrary, Wilson fermions completely solve the species
doubling problem, whereas it suffers from an explicit chiral symmetry breaking. The lattice simulation with Wilson fermions
 is more time-consuming than staggered fermions,  it can provide complementary information and crosscheck to simulations with other actions
 and establish a better understanding of QCD phase diagram.

In this paper, we attempt to investigate the RW transition line end points by lattice QCD with two
degenerate flavors of Wilson fermions. In Sec.~\ref{SectionLattice},
we define the lattice action with imaginary chemical potential and
the physical observables we calculate.  Our simulation results are
presented in Sec.~\ref{SectionMC} followed by discussions in
Sec.~\ref{SectionDiscussion}.

\begin{figure*}[t!]
\includegraphics*[width=0.49\textwidth]{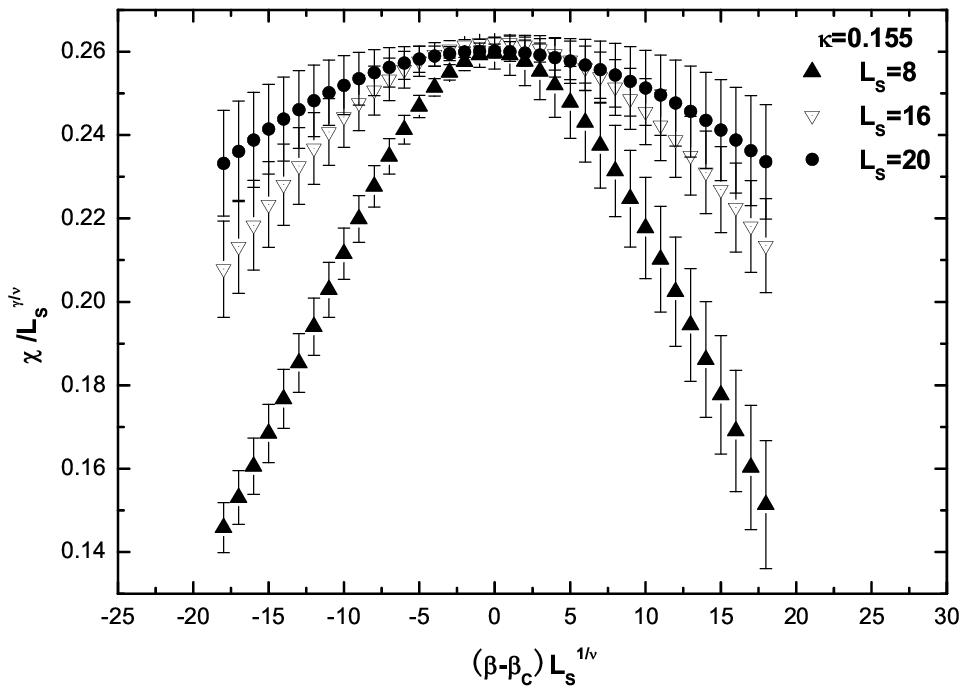}
\includegraphics*[width=0.49\textwidth]{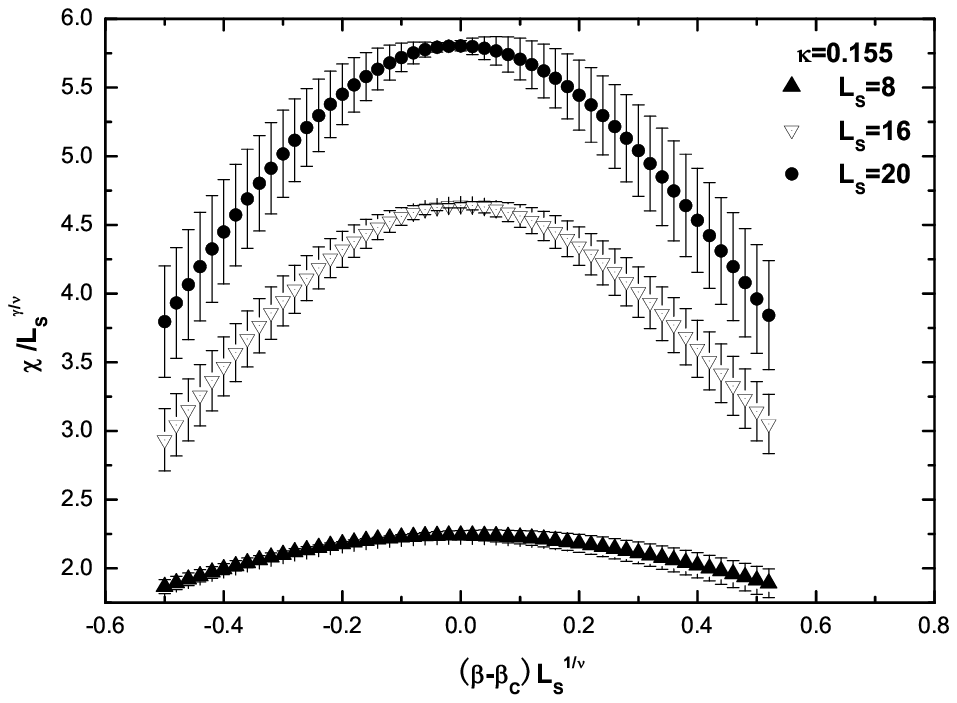}\\
\caption{\label{fig1} Scaling behavior of susceptbilities of the imaginary part of the Polyakov loop
according to the first order critical indexes (the left panel), and to the 3D Ising critical indexes (the  right panel) at $\kappa=0.155$.}
\end{figure*}

\begin{figure*}[t!]
\includegraphics*[width=0.49\textwidth]{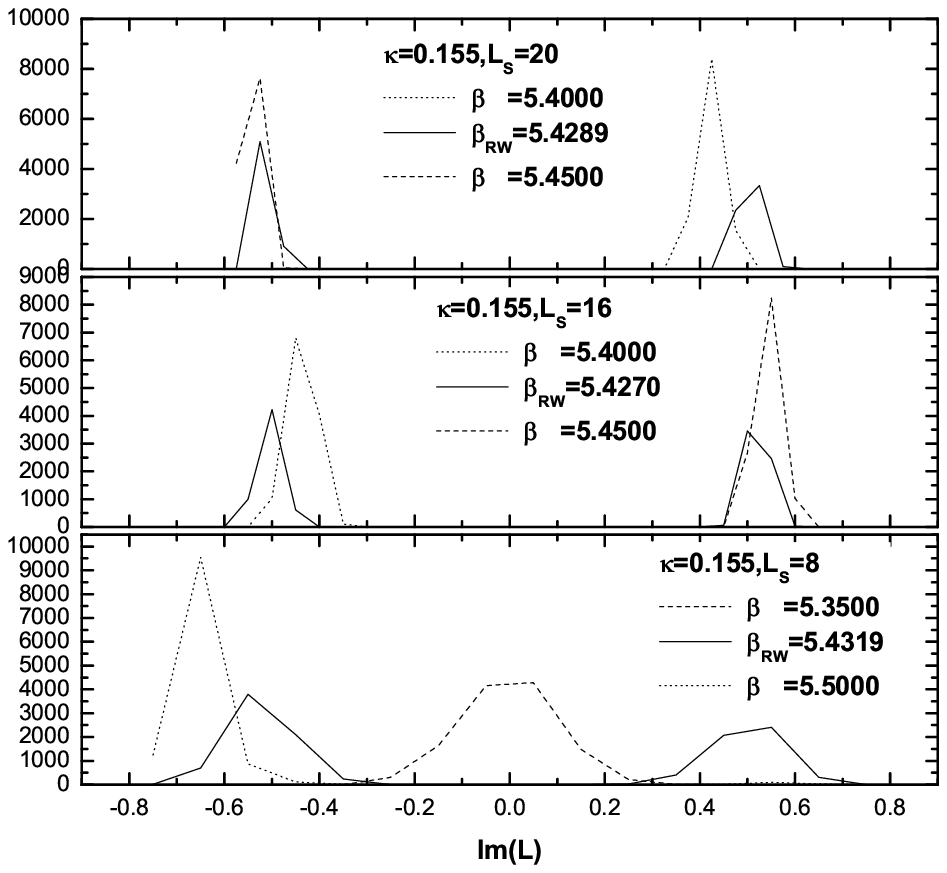}
\caption{Reweighted distributions of the imaginary part
of the Polyakov loop $Im(L)$  at the corresponding end point $\beta_{RW}$, and $\beta>\beta_{RW}$ and  $\beta<\beta_{RW}$ on  each lattice
spatial volume at $\kappa=0.155$.}
\label{fig051}
\end{figure*}

\begin{figure*}[t!]
\includegraphics*[width=0.49\textwidth]{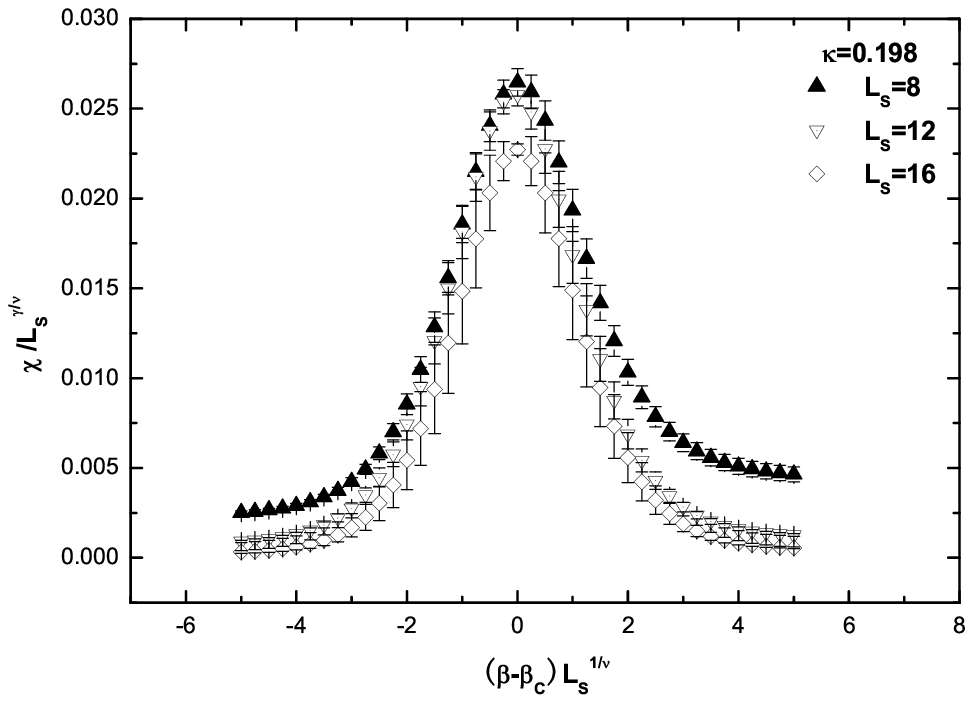}
\includegraphics*[width=0.49\textwidth]{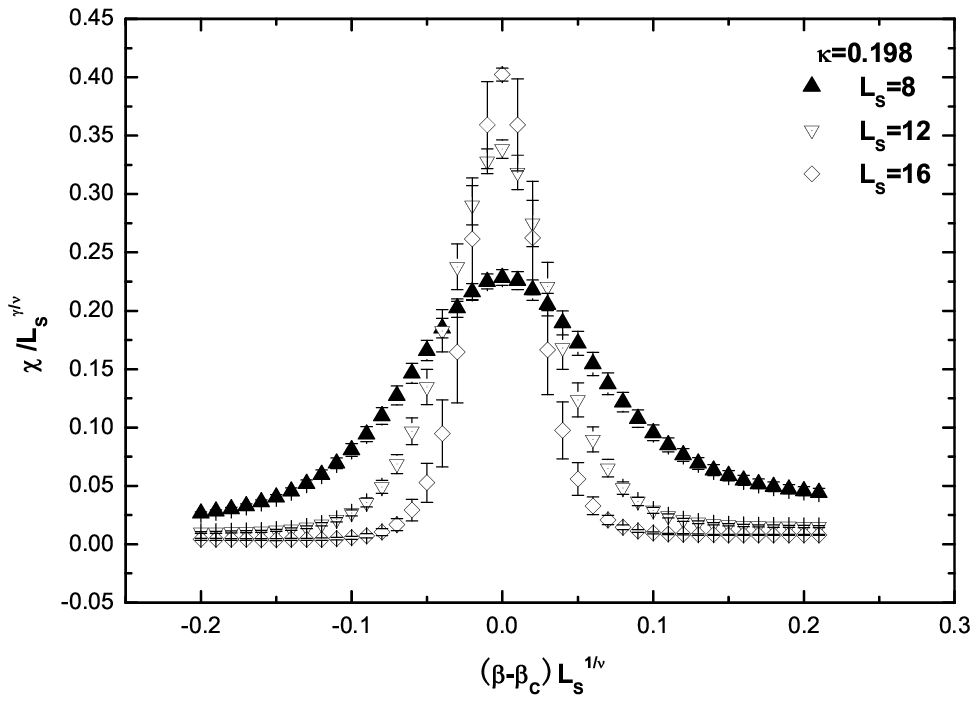}\\
\includegraphics*[width=0.49\textwidth]{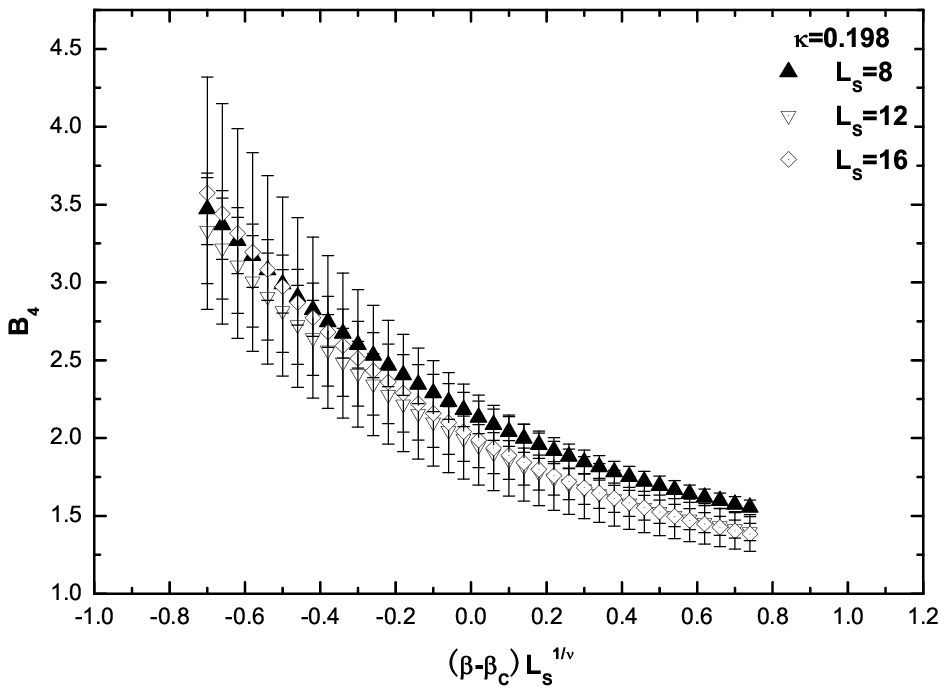}
\includegraphics*[width=0.49\textwidth]{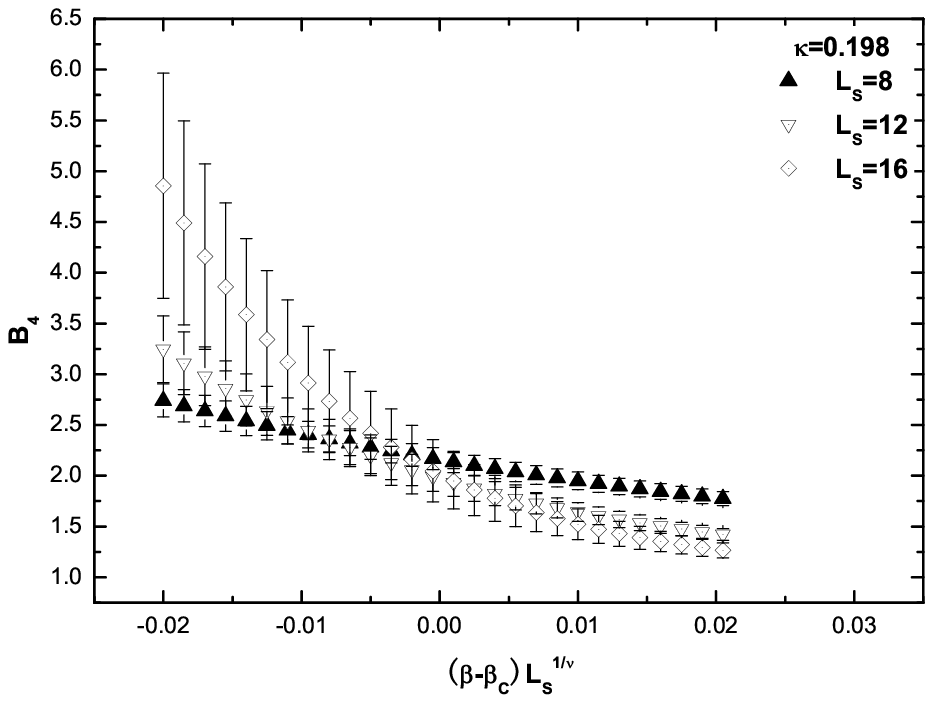}
\caption{\label{fig2} Scaling behavior of susceptibilities, and  Binder cumulants of the imaginary part of the Polyakov loop
according to the first order critical indexes (left panels), and to the 3D Ising critical indexes (right panels)at $\kappa=0.198$.}
\end{figure*}

\begin{figure*}[t!]
\includegraphics*[width=0.49\textwidth]{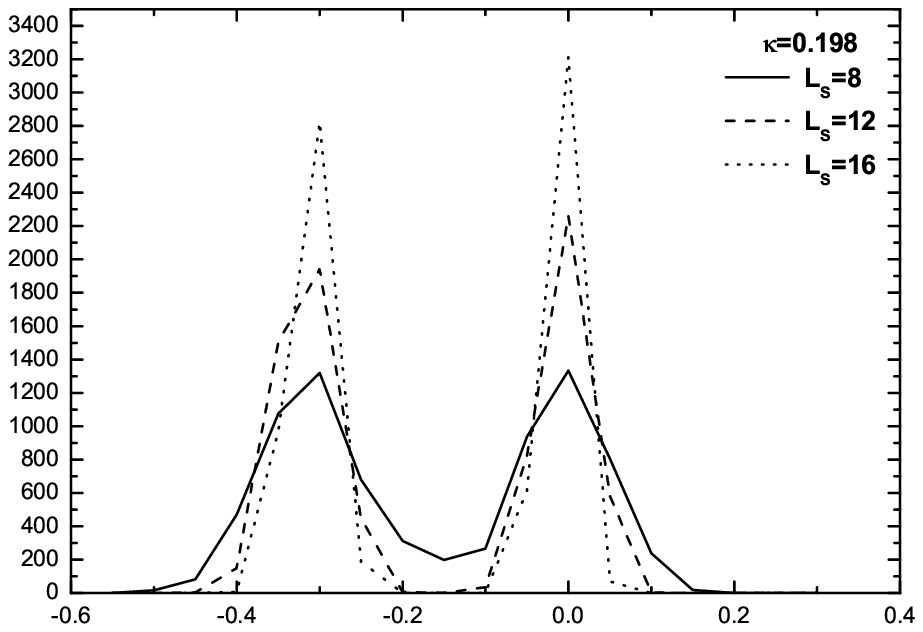}
\caption{Reweighted distributions of the imaginary part
of the Polyakov loop at $\kappa=0.198$ at the corresponding end points $\beta_{RW}$.}
\label{fig052}
\end{figure*}

\section{LATTICE FORMULATION WITH IMAGINARY CHEMICAL POTENTIAL}
\label{SectionLattice} We consider the partition function of system with
$N_f=2$ degenerate flavors of Wilson quarks with chemical potential on the
lattice
\begin{eqnarray}
\label{QCD_partition}
Z &= &\int [dU][d\bar\psi][d\psi]e^{-S_g-S_f} \nonumber \\
&=& \int [dU] \biggl({\rm Det} M [U,\theta]\biggr)^{N_f} e^{-S_g}.
\end{eqnarray}
where $S_g$ is the gauge action, and $S_f$ is the quark
action with the quark imaginary chemical potential $\mu_I=\theta T$.  For  $S_g$, we use the
standard one-plaquette action
\begin{eqnarray}
S_g=\beta\sum_p \biggl(1-\frac{1}{N}{\rm ReTr}U_p\biggr),
\end{eqnarray}
where $\beta=6/g^2$, and the plaquette variable $U_p$ is the ordered
product of link variables $U$ around an elementary plaquette. For
$S_f$, we use the the standard Wilson action
\begin{eqnarray}
S_f=\sum_{f=1}^{N_f}\sum_{x,y} {\bar
\psi}_f(x)M_{x,y}(U,\kappa,\mu){\psi}_f(y),
\end{eqnarray}
where $\kappa$ is the hopping parameter, related to the bare quark
mass $m$ and lattice spacing $a$ by $\kappa=1/(2am +8)$. The fermion
matrix is
\begin{eqnarray}
M_{x,y}(U,\kappa,\mu) & &  = {\delta_{x,y}} - \kappa \sum_{j=1}^{3}
\bigg[
(1-\gamma_{j})U_{j}(x)\delta_{x,y-\hat{j}} \nonumber \\
& &+ (1+\gamma_{j})U_{j}^{\dagger}(x-\hat{j})\delta_{x,y+\hat{j}}
\bigg]
\nonumber \\
&&- \kappa
   \bigg[(1-\gamma_{4})e^{a\mu}U_{4}(x)\delta_{x,y-\hat{4}}
\nonumber \\
&&+
   (1+\gamma_{4})e^{-a\mu}U_{4}^{\dagger}(x-\hat{4})\delta_{x,y+\hat{4}}\bigg].
\end{eqnarray}

We carry out simulations at $\theta=\pi$. As it is pointed out that the system is invariant under the charge
 conjugation at $\theta =0,\pi$, when $\theta $ is fixed~\cite{Kouno:2009bm}. But the $\theta$-odd quantity $O(\theta)$ is not invariant at $\theta=\pi$
 under charge conjugation. When $T < T_{RW}$, $O(\theta)$ is a smooth function of $\theta$, so it is zero at $\theta=\pi$.
 Whereas when $T > T_{RW}$,  the two charge violating solutions cross each other at $\theta=\pi$. Thus the charge symmetry is
 spontaneously broken there and the $\theta$-odd quantity $O(\theta)$ can be taken as order parameter . In this paper, we take the imaginary part of Polyakov loop as the order parameter.

The Polyakov loop $ L $ is defined as the following:
\begin{eqnarray}
\langle L \rangle=\left\langle \frac{1}{V}\sum_{\bf x}{\rm  Tr}
\left[ \prod_{t=1}^{N_t} U_4({\bf x},t) \right] \right\rangle ,
\end{eqnarray}
here and in the following, $V$ is the spatial lattice volume. To simplify the notations, we use $X$ to represent
the imaginary part of Polyakov loop $L$, $X={\rm Im}(L)$.

The susceptibility of imaginary part of Polyakov loop $\chi$ is defined as
\begin{eqnarray}
\chi= V \left\langle( X - \langle  X\rangle)^2\right\rangle ,
\end{eqnarray}
which is expected to scale as:~\cite{D'Elia:2009qz,Bonati:2010gi}
\begin{eqnarray}\label{chi_scaling}
\chi= L_s^{\gamma/\nu}\phi(\tau L_s^{1/\nu}) ,
\end{eqnarray}
where $\tau $ is the reduced temperature $\tau=(T-T_{RW})/T_{RW}$, $V=L_s^3$ .
This means that the curves $\chi/L_s^{\gamma/\nu}$ at different lattice volume should collapse with the same curve when plotted against $\tau L_s ^{1/\nu}$.
In the following, we employ $\beta-\beta_{RW}$ in place of $\tau=(T-T_{RW})/T_{RW}$. The critical exponents relevant to our study
are collected in Table.~\ref{critical_exponents}~\cite{Bonati:2010gi,Pelissetto:2000ek}.
\begin{table}[htmp]
\begin{center}
\begin{tabular}{|c|ccc|}
\hline
          &  $\nu$     & $\gamma $    &       $\gamma/\nu$    \\
 \hline
3D ising  & 0.6301(4)  & 1.2372(5)    &        1.963                  \\
tricritical & 1/2      &  1           &           2                         \\
 first order    & 1/3  &  1           &               3                        \\
 \hline
\end{tabular}
\end{center}
\caption{\label{critical_exponents}Critical exponents relevant to our study.}
\end{table}

We also consider the Binder cumulant of the imaginary part of Polyakov loop which is defined as the following:
\begin{eqnarray}\label{binder_scaling}
B_4=\left\langle ( X - \langle  X\rangle)^4\right\rangle /
    \left\langle ( X - \langle  X\rangle)^2\right\rangle^2 ,
\end{eqnarray}
with  $\langle X \rangle =0$. In the thermodynamic limit, $B_4(\beta)$
takes on the values 3,  1.5, 1.604, 2 for crossover, first order triple point, 3D Ising and tricritical transitions, respectively.
However, on finite spatial volumes, the steps are smeared out to continuous functions. In the vicinity of the RW transition line
end points, $B_4$ is a function of $x=(\beta-\beta_{RW})L_s^{1/\nu}$ and can be expanded as a series~\cite{deForcrand:2010he,Philipsen:2010rq,Bonati:2012pe}.
\begin{eqnarray}\label{binder_scaling_02}
B_4=B_4(\beta_c,\infty)+a_1x+a_2x^2+\cdots,
\end{eqnarray}

\section{MC SIMULATION RESULTS}
\label{SectionMC}
In this section, we will present our results for simulating QCD with
two  degenerate flavors of Wilson fermions at finite temperature $T$
and imaginary chemical potential $i\mu_I$. Both the $\phi$ algorithm with a Metropolis accept/reject
step and the $R$
algorithm are used~\cite{Gottlieb:PRD:35:3972}.    The simulations are performed on lattice with different spatial volume
with temporal extent $N_t=4$
 at $\kappa=0.155,\,0.160,\,0.165,\,0.168,\,0.170,\,0.175,\,0.180,\,0.190,\,0.198 $. For each $\kappa$ value, we carry out simulations
 on lattice of size $L_s=8,\, 12,\, 16$, and for some $\kappa $ values, lattice of size $L_s=10$ or/and $L_s=20$ are also used.
 Simulations are carried out with  $\phi$ algorithm with a Metropolis accept/reject
step with the acceptance rate ranging from $42-93\% $,  The other simulations are carried out in terms of $R$
algorithm with the molecular dynamics time step  $\delta\tau=0.01$.
Ref.~\cite{Gottlieb:PRD:35:3972} pointed out that
R-algorithm has errors of order $O(\delta \tau^2)$, so the correct
results of this algorithm consists of extrapolation to zero
stepsize.  However, in practice a short-cut without extrapolation is
used. Recently, the exact RHMC algorithm is invented which
 also allows many improvements~\cite{Clark:2002vz}. In our simulation, $\delta \tau =0.01$ is sufficiently smaller
compared with the statistical errors of our simulations.
 There are 20 molecular steps  for each trajectory.
We generate 20,000 trajectories after 10,000 trajectories as warmup.
Ten trajectories are carried out between measurements.  We use
the conjugate gradient method to evaluate the fermion matrix
inversion.

On each lattice size, we make simulations at typically 4-6 different $\beta$ values. For fixed $i\mu_I=i\pi T$, there is transition in $T$
between  the low temperature phase  and the high temperature phase.
In order to determine the RW transition line end point $\beta_{RW}$ from the peak of susceptibilities, we use the data obtained through simulations at the 4-6 different $\beta$ values,
and calculate  susceptibilities at additional $\beta$ values,
by employing  the Ferrenberg-Swendsen reweighting
method~\cite{Ferrenberg:1989ui}.

Let us first present the critical couplings $\beta_{RW}$ on different spatial volume at different $\kappa$ in Table.~\ref{critical_beta}.
\begin{table}[htmp]
\caption{\label{critical_beta}Results of critical couplings $\beta_{RW}$ on different spatial volume at different $\kappa$, we also make simulations on lattice $8^3\times4$ at $\kappa=0.185,\, 0.195$, the critical couplings $\beta_{RW}$ are $4.8810(20),\,4.6610(20)$, respectively.}
\begin{ruledtabular}
\begin{center}
\begin{tabular}{c|ccccc}

$\kappa$    &     $8 $       &      $ 10  $   &  $  12   $       &    $16 $       & $  20 $       \\  \hline
0.155       & $5.4319(40)$   &                &  $5.3887(40)$    & $5.427(10)$    & $5.4289(50)$  \\
0.160       & $5.361(50) $   &  $5.365(30) $  &  $ 5.347(10) $   & $5.3499(60) $  &               \\
0.165       & $5.2566(90)$   &  $5.262(13) $  &  $ 5.2493(20)$   & $5.2412(10)$   & $5.2581(10)$  \\
0.168       & $5.206(15) $   &                &   $5.2103(22)$   & $ 5.2167(6)$   & $5.2181(10)$  \\
0.170       & $5.1645(50)$   &                &   $5.1722(10)$   & $ 5.1770(5)$   & $5.1785(2)$   \\
0.175       & $5.0781(30)$   &                &   $5.0838(50)$   & $ 5.0882(40)$  & $5.1095(30)$   \\
0.180       & $4.9802(20)$   &                &   $5.0388(60)$   & $ 5.0391(40)$  &               \\
0.190       & $4.7800(20)$   &                &   $4.7658(10)$   & $ 4.7883(3)$   &               \\
0.198       & $4.5910(20)$   &                &   $4.5955(10)$   & $ 4.5980(2)$   &               \\
\end{tabular}
\end{center}
\end{ruledtabular}
\end{table}

The presence of a first order phase transition at the end point of Roberge-Weiss transition line at $\kappa=0.155$ can be
found from the scaling behavior of the susceptibilities of the imaginary part of Polyakov loop $\chi$ presented in
Fig.~\ref{fig1}.  From Fig.~\ref{fig1} we can find that the rescaling quantities
$\chi/{L_s^{\gamma/\nu}}$
plotted against $(\beta-\beta_{RW})L_s^{1/\nu}$ does not fall on the same curve completely, whereas peaks of the rescaling quantities $\chi/{L_s^{\gamma/\nu}}$ obviously exhibit scaling behavior which conforms to the first order transition.
From Eq.~(\ref{chi_scaling}),  we can find that the index $\gamma/\nu$ regulates the height of peaks while the
index $\nu$ regulates the width of peaks. As a comparison, we also present
the behavior according to the 3D Ising transition index in the right panel of Fig.~\ref{fig1} from which we can find that
large deviation from the 3D Ising scaling behavior manifest clearly. At $\kappa=0.160$, similar observations of susceptibilities
as those at $\kappa=0.155$ can be found.

In Fig.~\ref{fig051}, we present  reweighted distributions of the imaginary part of Polyakov loop $Im(L)$ at the corresponding $\beta_{RW} $ and two $\beta$ values on lattice size $L_s=8,\,16,\,20$. On each lattice size, at $\beta_{RW}$,  reweighted distribution of $Im(L)$  exhibits two-state signal, while, at $\beta>\beta_{RW}$ and $\beta<\beta_{RW}$,
reweighted distributions of $Im(L)$  do not exhibit two-state signal. At other $\kappa$ values, reweighted distributions of the imaginary part of Polyakov loop $Im(L)$ at the corresponding $\beta_{RW} $, $\beta>\beta_{RW}$ and $\beta<\beta_{RW}$ on each lattice size
have the same observations as those at $\kappa=0.155$. For clarity, we only present the result at $\kappa=0.168$ in the following.

We also make simulations at $\kappa=0.190,\,0.198$, the results of simulations at $\kappa=0.198$ are presented in Fig.~\ref{fig2}, and Fig.~\ref{fig052}. From the two upper panels of
Fig.~\ref{fig2}, we can find that the first order transition indexes are more suitable to describe the behavior than the
3D Ising ones. This situation can be made  clearer when we look at the $B_4$ behavior depicted in down panels of Fig.~\ref{fig2} from which we can find that the quantities of Binder cumulant plotted against  rescaling $\beta$ fall on the same curve completely. Note that from Eq.~(\ref{binder_scaling}), the scaling behavior of Binder cumulants is governed by the critical index $\nu$ which also determines the width of peaks of the rescaling quantities $\chi/{L_s^{\gamma/\nu}}$. The fact that the value of $\nu$ for first order transition accounts for the width of peaks of  $\chi/{L_s^{\gamma/\nu}}$ better than the second order transition show that the transition is first order,  and this situation is confirmed by the scaling behavior of Binder cumulant $B_4$. We also present reweighted distribution of the imaginary part of Polyakov loop at $\beta_{RW}$ at $\kappa=0.198$ in Fig.~\ref{fig052} which exhibits two-state signal. At $\kappa=0.190$, similar observations as those at $\kappa=0.198$ can be observed.

\begin{figure*}[t!]
\includegraphics*[width=0.49\textwidth]{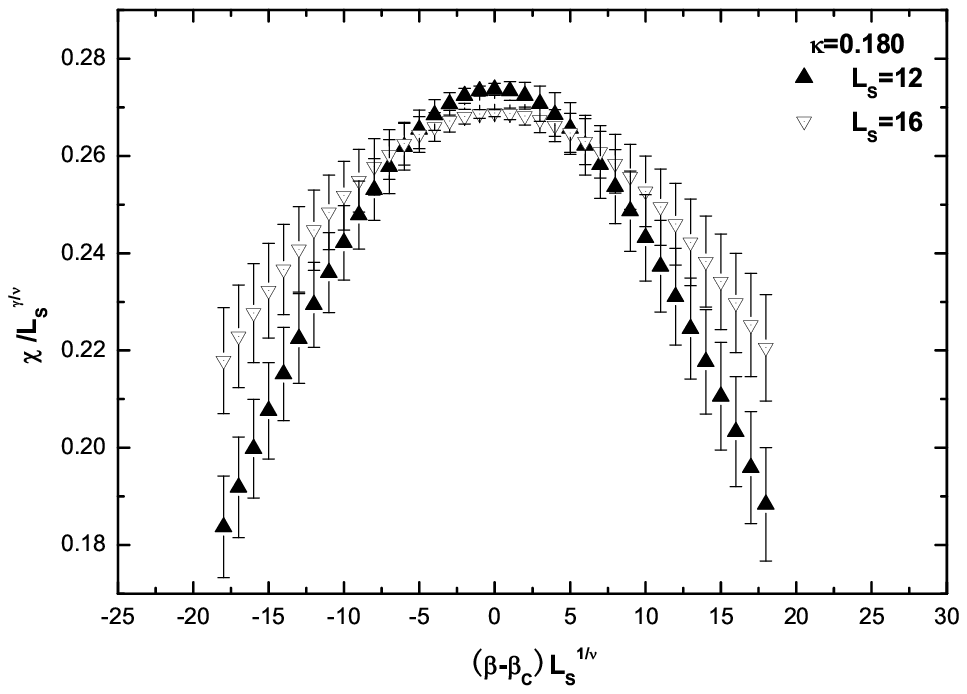}
\includegraphics*[width=0.49\textwidth]{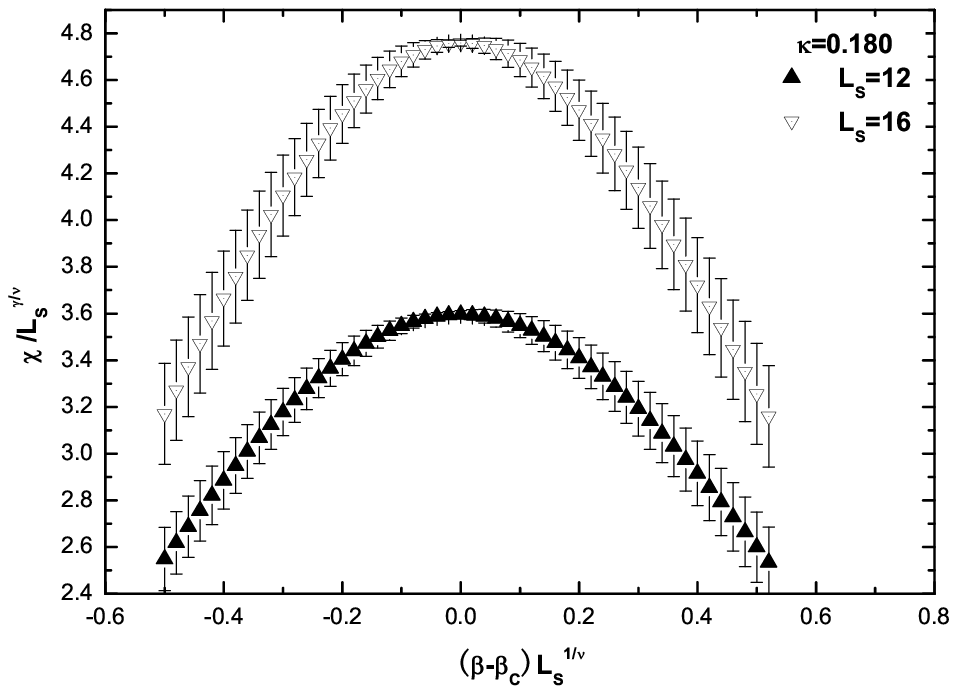}\\
\includegraphics*[width=0.49\textwidth]{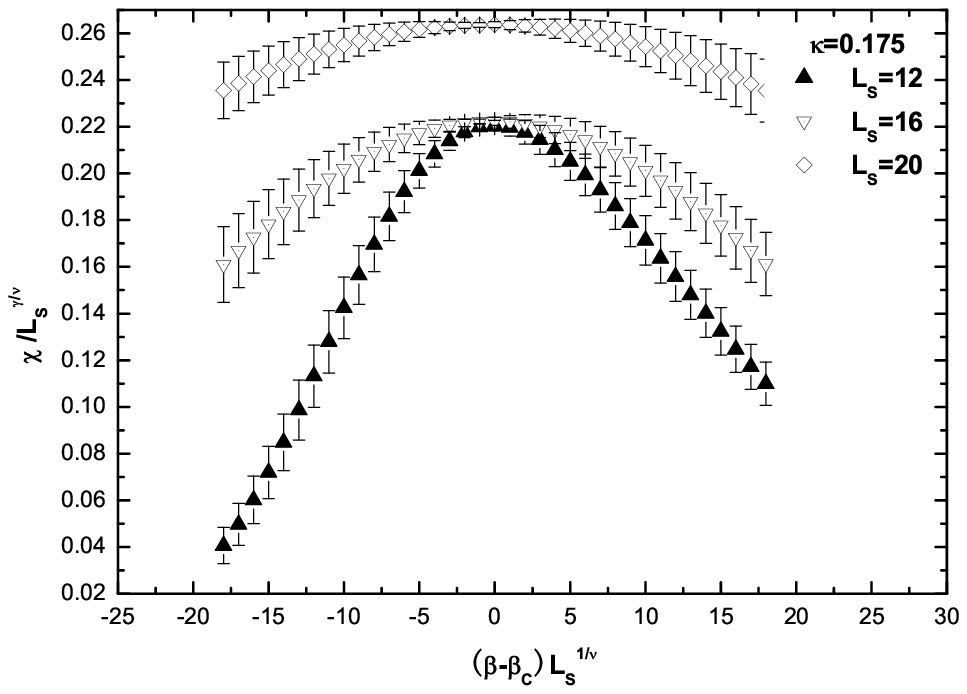}
\includegraphics*[width=0.49\textwidth]{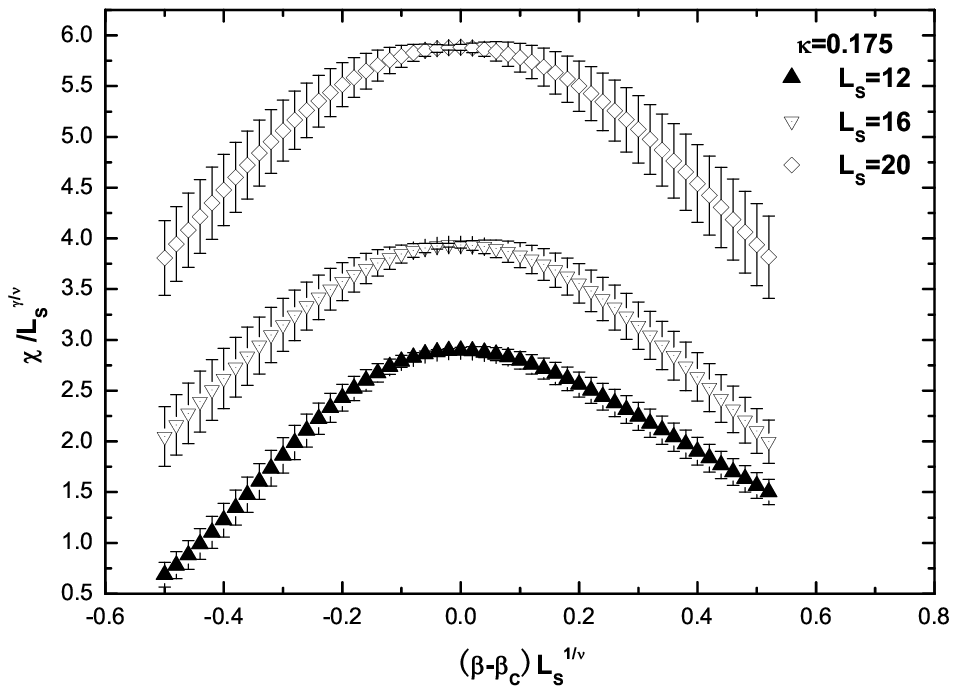}\\
\includegraphics*[width=0.49\textwidth]{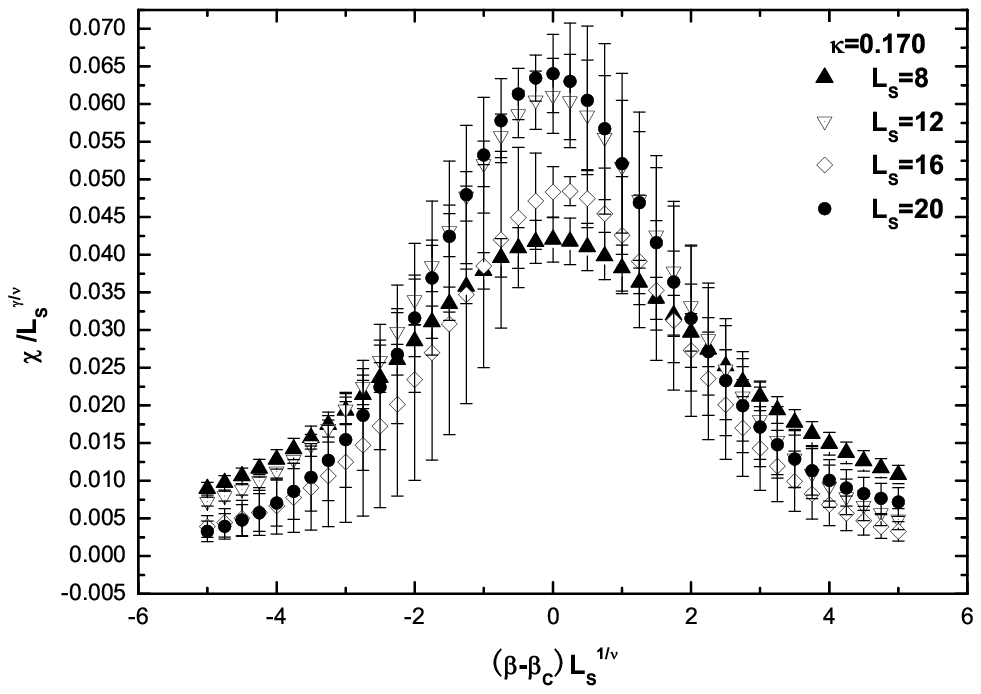}
\includegraphics*[width=0.49\textwidth]{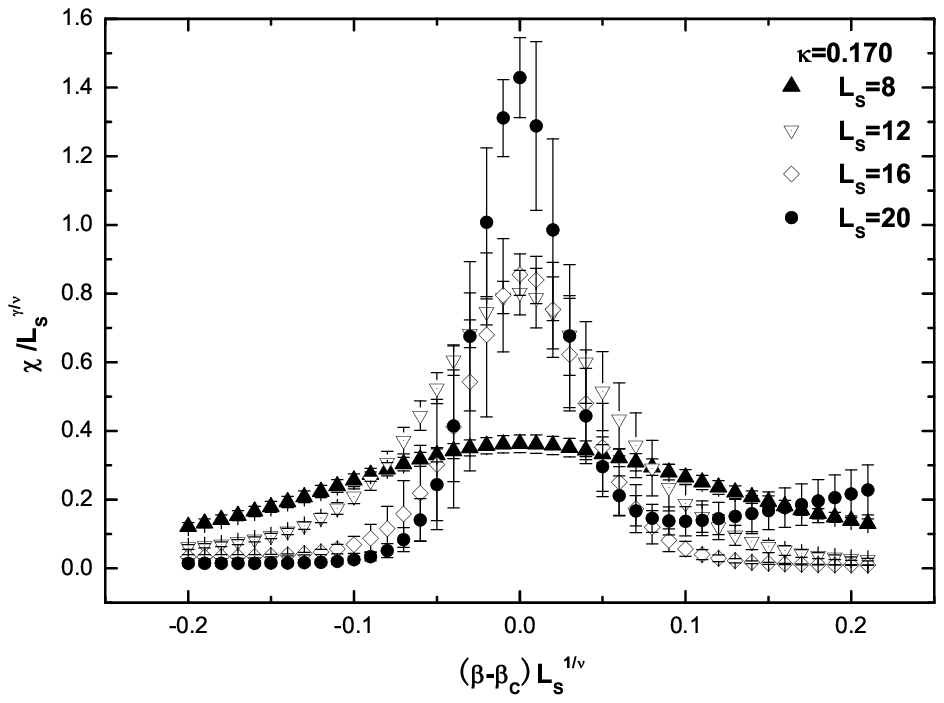}
\caption{Scaling behavior of susceptibilities of the imaginary part
of the Polyakov loop according to first order critical indexes (left panels) and to the 3D Ising critical indexes (right panels).}
\label{fig4}
\end{figure*}

\begin{figure*}[t!]
\includegraphics*[width=0.49\textwidth]{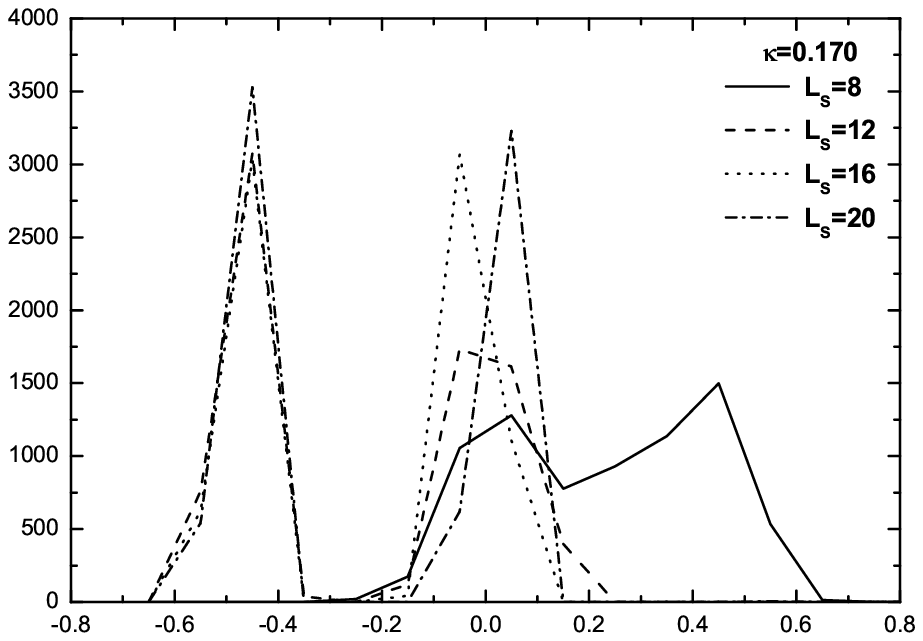}
\includegraphics*[width=0.49\textwidth]{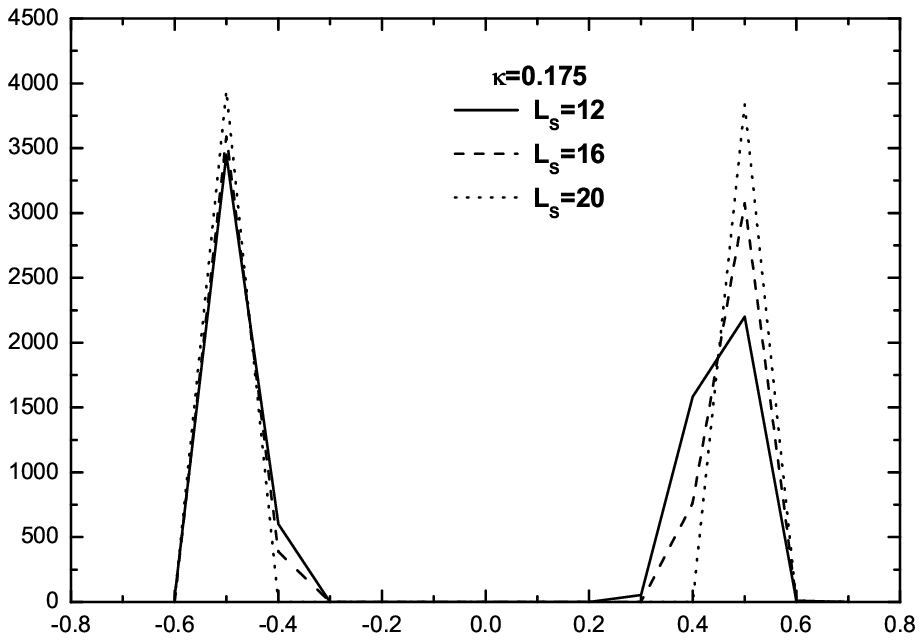}\\
\includegraphics*[width=0.49\textwidth]{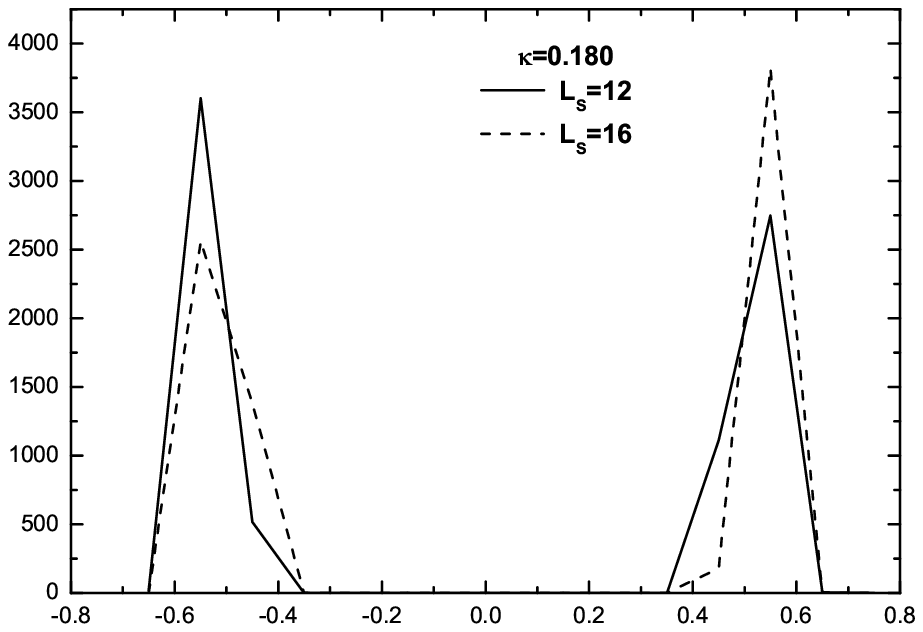}
\caption{Reweighted distributions of the imaginary part
of the Polyakov loop at $\kappa=0.170,\,0.175,\,0.180$ at the corresponding end points $\beta_{RW}$.}
\label{fig053}
\end{figure*}

The results of simulations at $\kappa=0.170, \, 0.175,\,0.180$ are shown in Fig.~\ref{fig4}.  In view of  the fact that large finite-size corrections are observed in simple spin models even when the transition is first order~\cite{deForcrand:2010he,Billoire:1992ke},  we can find that the first order transition indexes perform much better than  the second order transition ones. This observation can be enhanced from the  reweighted distribution of the imaginary part of Polyakov loop presented in Fig.~\ref{fig053}.

\begin{figure*}[t!]
\includegraphics*[width=0.49\textwidth]{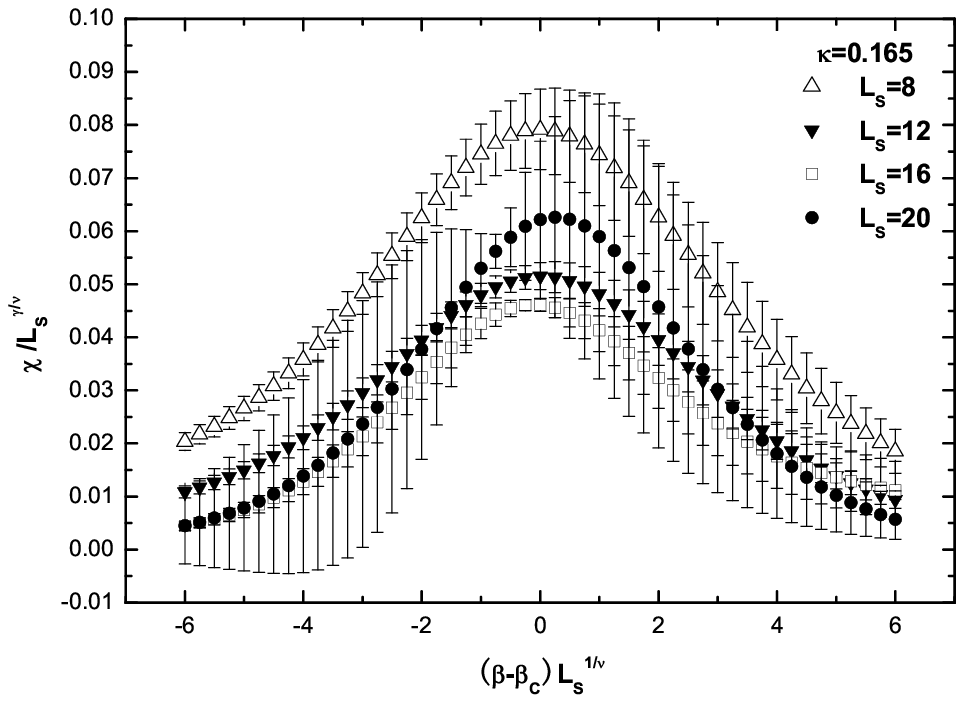}
\includegraphics*[width=0.49\textwidth]{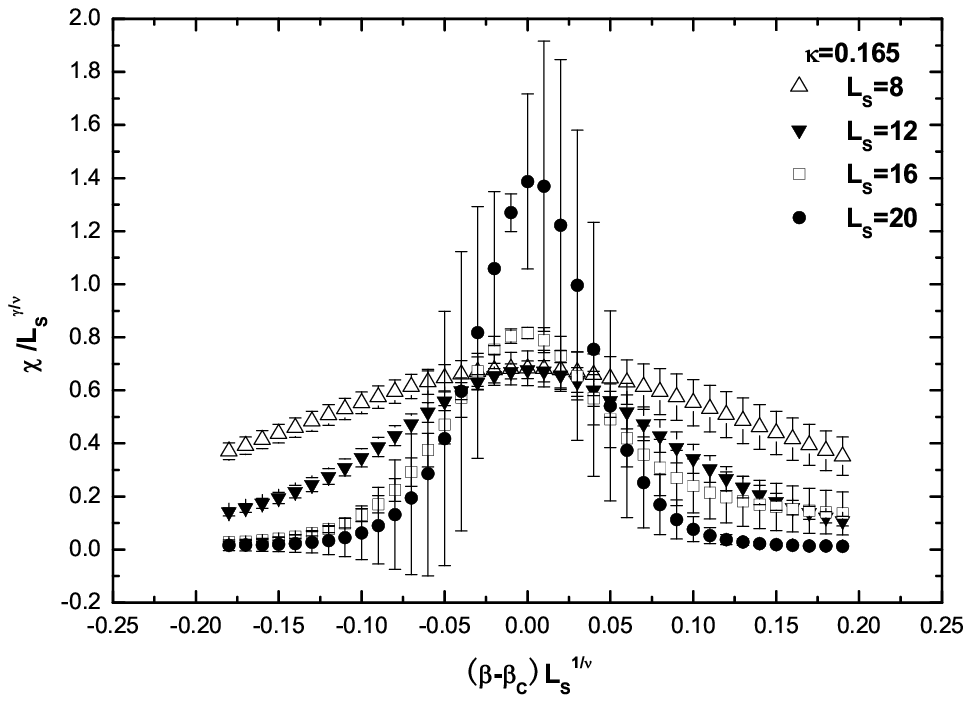}\\
\includegraphics*[width=0.49\textwidth]{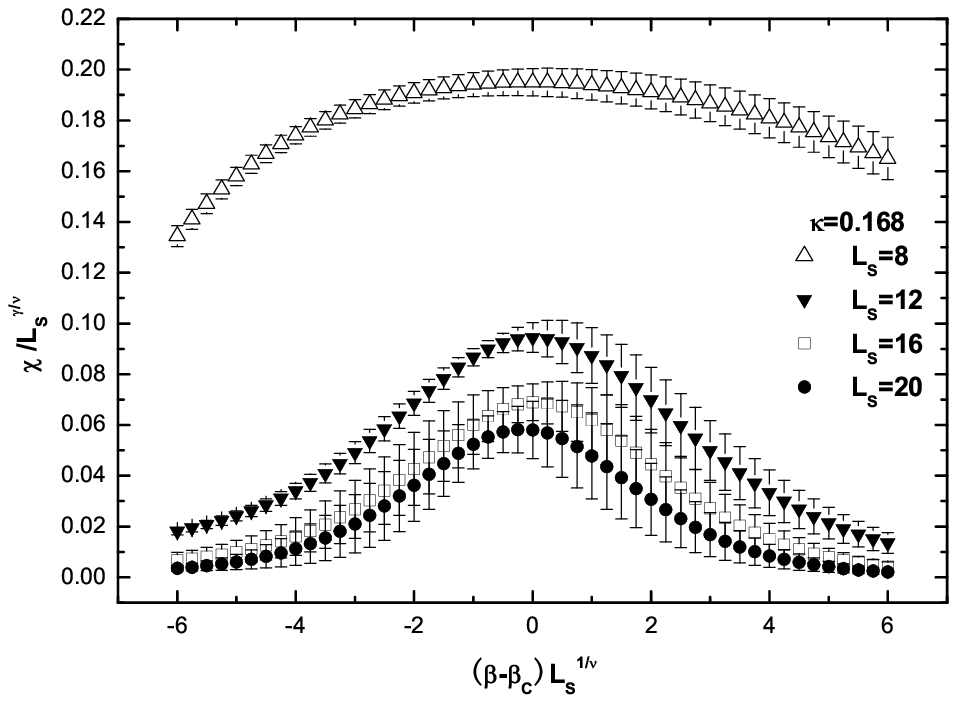}
\includegraphics*[width=0.49\textwidth]{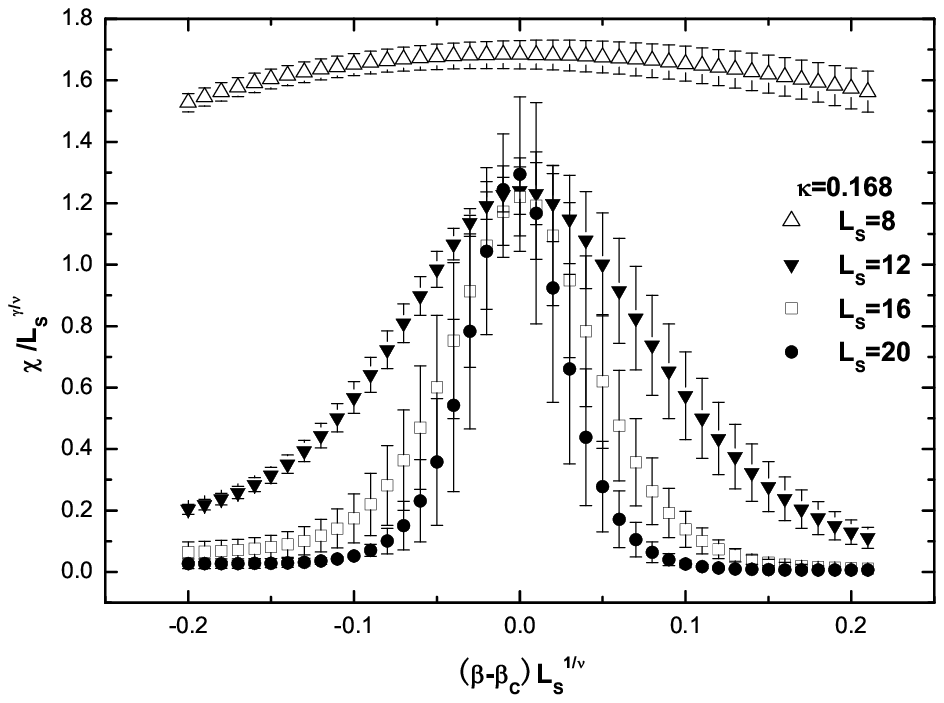}
\caption{Scaling behavior of  susceptibilities of the imaginary part
of the Polyakov loop according to the first order critical indexes (left panels)
and to the 3D Ising critical indexes (right panels).}
\label{fig5}
\end{figure*}

\begin{figure*}[t!]
\includegraphics*[width=0.49\textwidth]{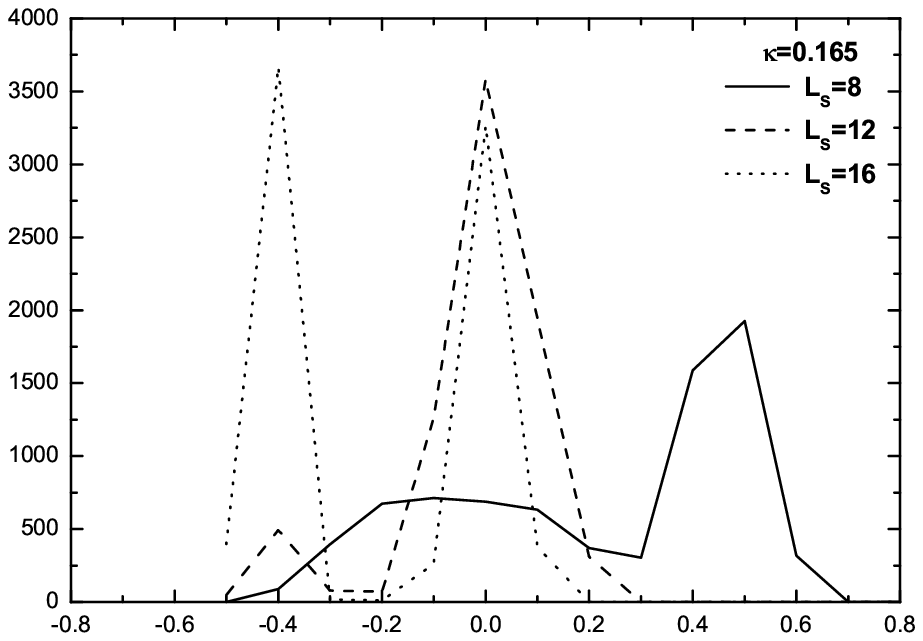}
\caption{Reweighted distributions of the imaginary part
of the Polyakov loop at $\kappa=0.165$ at the corresponding end points $\beta_{RW}$.}
\label{fig054}
\end{figure*}

\begin{figure*}[t!]
\includegraphics*[width=0.49\textwidth]{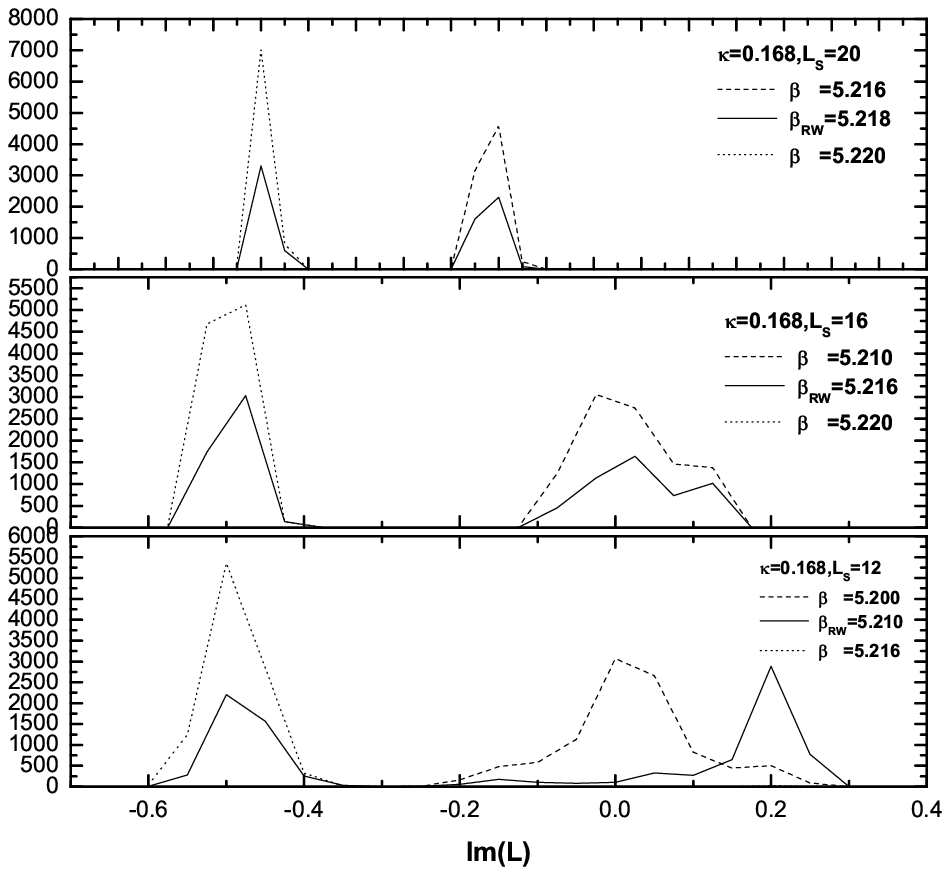}
\caption{Reweighted distributions of the imaginary part
of the Polyakov loop $Im(L)$  at the corresponding end point $\beta_{RW}$, and $\beta>\beta_{RW}$ and  $\beta<\beta_{RW}$ on  each lattice spatial volume at $\kappa=0.168$.}
\label{fig055}
\end{figure*}

Comparing to the above results, it is difficult to determine the nature of RW transition line end points at $\kappa=0.165,\,0.168$ results of which
are presented in Fig.~\ref{fig5}. However, when we look at the behavior at large lattice size presented in Fig.~\ref{fig5}, it is a reasonable conclusion that the behavior of RW transition line end points at $\kappa=0.165,\,0.168$ are of first order. This conclusion can be enhanced  when we look at the reweighted distributions of $Im(L)$ at the end point $\beta_{RW}$ at $\kappa=0.165$ presented in Fig.~\ref{fig054}. and
reweighted distributions of  $Im(L)$ at the corresponding $\beta_{RW} $, $\beta>\beta_{RW}$ and $\beta<\beta_{RW}$ on  lattice size
$L_S=12,\,16,\,20$ at $\kappa=0.168$ presented in Fig.~\ref{fig055}.

\section{DISCUSSIONS}\label{SectionDiscussion}

We have studied the nature of critical end points of Roberge-Weiss transition of two flavor lattice QCD with
Wilson fermions. When $\mu = i\pi T$, the imaginary part of Polayakov loop is the order parameter for studying
the transition from low temperature phase to high temperature one.
Within the imaginary chemical potential formulation, the partition function is periodic in
imaginary chemical potential. The different Z(3) sectors are
characterized by the phase of Polyakov loop. The Roberge-Weiss transition
which occurs at $\mu_I/T=2\pi(k+1/2)/3$ is of first order in the
high temperature phase, whereas it is of crossover in the low temperature phase.

Our simulations are carried out at 9  values of $\kappa$ on different 3-4 spatial volumes.  Our lattice $N_t=4 $ is coarse. In Ref.~\cite{Bitar:1993xd}, the lattice spacing
 with 2 flavor Wilson fermions at $\beta=5.3$ is estimated to be $ 0.12-0.13 \ {\rm fm}$.
 In Ref.~\cite{Iwasaki:1996zt}, the lattice spacing with  2 flavor Wilson fermions  is estimated to be $ 0.246 \ {\rm fm}$ which is found almost independent of $\beta$ in the range of $\beta=3.0-4.7$. In our simulations, $\beta$ varies roughly from $4.6$ to $5.4$,  thus, the lattice spacing $a$ is estimated to be $a \sim 0.12-0.25 \ {\rm fm}$.

In order to estimate the pseudo-scalar meson mass $m_\pi$, the vector meson
mass $m_\rho$ and ratios $m_\pi/m_\rho,\,T_c/m_\rho$ at our simulation
points, we use the data in Table II in Ref.~\cite{Bitar:1990si}. By using the
standard quark and gauge action,  Bitar {\it et al}. studied hadron thermodynamics with Wilson fermions on lattice $8^3 \times 4$
and calculated the zero temperature hadron mass on lattice $8^3 \times 16$ with dynamical fermions. We compile their  results and present in the following:
at $\kappa=0.16,\,\beta=5.28$, $m_\pi/m_\rho= 0.943(3),\,T_c/m_\rho=0.19425(7)$, at $\kappa=0.17,\,\beta=5.12$,
$m_\pi/m_\rho= 0.899(4),\,T_c/m_\rho=0.2066(8)$,  at $\kappa=0.18,\,\beta=4.94$,
$m_\pi/m_\rho= 0.836(5),\,T_c/m_\rho=0.224(1)$, and at $\kappa=0.19,\,\beta=4.76$,
$m_\pi/m_\rho= 0.708(7),\,T_c/m_\rho=0.245(2)$. Using the lattice spacing estimated in the above, we find that at $\kappa=0.190,\,\beta=4.76$, $m_\pi=578(2)\ {\rm MeV}$, at $\kappa=0.160,\,\beta=5.28$, $m_\pi=1991(7)\ {\rm MeV}$. Comparing the values of $\kappa,\, \beta$ at simulation points in Ref.~\cite{Bitar:1990si}
with ours. we can roughly estimate the pseudo-scalar meson mass $m_\pi$. Using the estimated lattice spacing, we can estimate that Roberge-Weiss transition point temperature varies from $197-410 \ {\rm MeV}$ in our simualtions.

We consider
 the peak behaviour, reweighted distribution and  Binder cumulant of order parameter around the critical end point $\beta_{RW}$,
At $\kappa=0.190, \, 0.198 $, the three observables' behaviour show that transition at the end point is of first order which means the
end point is a triple point.  At
  $\kappa=0.155\, ,0.160 $, the peak behavior at the end point are more consistent with that of transition of a triple point than
that of 3D Ising transition behaviour. Similar observations can be observed at $\kappa=0.170, \,0.175,\, 0.180$.

At $\kappa=0.165,\,0.168$, it becomes difficult to discern the peak behavior between 3D Ising transition class and  triple point,
however, when we look at the peak behavior at large lattice size, it is a reasonable conclusion that the behavior of RW transition line end points are of first order. This conclusion is enhanced by the reweighted distribution of  order parameter.

 We also fit  Eq.~(\ref{binder_scaling_02}) to the calculated Binder cumulant data to
 extract the value of critical index $\nu$. At $\kappa=0.165,\,0.168$, $\nu=0.3661,\,0.3594$, respectively, and these values  conform to first order transition.

In Ref.~\cite{Aarts:2010ky}, the locations of triple points are determined.
In Ref.~\cite{D'Elia:2009qz,Bonati:2010gi} and Ref.~\cite{deForcrand:2010he}, the simulations with staggered fermions show that  phase diagram of two flavor and three flavor QCD at imaginary
chemical potential $\mu=i\pi T$ are characterized by two tricritical points, respectively. Our simulations have no  evidence that shows
 the existence of tricritical points separating second order region from the first order region.  Considering these results, our investigation requires further extensive numerical simulations  which extend to a larger range of quark mass region. This work is under progress.

\begin{acknowledgments}
 We thank the referee for the comments very much. We modify the MILC collaboration's
public code\cite{Milc} to simulate the theory at imaginary chemical
potential. This work is supported by the National Science Foundation of China (NSFC) under Grants No. 11105033. The work was carried out at National Supercomputer Center in Tianjin,and the calculations were performed on TianHe-1A.
\end{acknowledgments}

\end{document}